\documentclass[adp,fleqn]{w-art}
\usepackage{graphicx}
\usepackage{epstopdf}
\usepackage[dvips]{epsfig}
\begin{document}
           \csname @twocolumnfalse\endcsname

\title{On the quantum statistics of bound states within the
Rutherford model of matter}
\author[Ebeling]{W. Ebeling$^1$\footnote{Corresponding
     author: e-mail: {\sf ebeling@physik.hu-berlin.de}}}
\address[\inst{1}]{Institut f\"ur Physik, Humboldt-Universit\"at zu Berlin,
Newtonstr. 15, D-12489 Berlin}
\author[Kraeft]{W.D. Kraeft$^{2,3}$\footnote{e-mail: {\sf
wolf-dietrich.kraeft@uni-rostock.de}}}
     \address[\inst{2}]{Institut f\"ur Physik, Ernst-Moritz-Arndt-Universit\"at Greifswald,
Felix-Hausdorff-Str.6, D-17487 Greifswald}
\author[R\"opke]{G. R\"opke$^3$\footnote{e-mail: {\sf
gerd.roepke@uni-rostock.de}}}
     \address[\inst{3}]{Institut f\"ur Physik, Universit\"at Rostock,
Universit\"atsplatz 3, D-18055 Rostock}
\begin{abstract}
The quantum statistical treatment of the Rutherford model, considering matter as a system of
point charges (electrons and nuclei) is analyzed. First, in the historical
context, the solutions of different fundamental problems, such as the
divergence of the partition function, elaborated by Herzfeld, Planck,
Brillouin and Rompe - most of the relevant papers published in the Annalen der
Physik, are discussed. Beyond this, the modern state
of art is presented and new results are given which explain, why bound states according to a discrete part of the spectra
occur only in a valley in the temperature-density plane.
Based on the actual state of the quantum statistics of Coulomb systems,
virial expansions within the canonical ensemble and the grand ensemble and combinations are derived.
The following transitions along isotherms are studied:
(i) the formation of bound states occurring by increasing the density from low to moderate values,
(ii) the disappearance of bound state effects at higher densities due to medium effects. Within the physical picture we calculate isotherms of pressure for Hydrogen
in a broad density region and show that in the region between $20\, 000$ K and $100\, 000$ K and particle densities below $10^{22}$ cm$^{-3}$ the cross-over from full to partial ionization may be well described by the contributions of extended ring diagrams
and ladder diagrams.
\end{abstract}

\date{Version 13.12.2011 }

\maketitle

\section{Introduction}

Hundred years ago, Rutherford invented a new model for the interpretation of
existing scattering experiments on the scattering of electrons on matter. In May
1911, Rutherford came forth with a model for the structure of atoms which
provided a first understanding why electron scattering can so deeply penetrate
into the interior of atoms, so far unexpected experimental results
\cite{Rutherford}.
Rutherford explained his scattering results as the passage of a high speed electron through an atom having a
positive central charge  $+N e$, and surrounded by a compensating charge of $N$ electrons \cite{Rutherford}.
In Rutherford's model the atom is nearly empty, it is made up of a central charge and electrons and the
glue keeping the system together are the Coulomb forces. Today we denote the positive central charge
as the atomic nucleus, though Rutherford did not use the term "nucleus" in his paper.
The central point charge is surround by a cloud of point electrons orbiting around the nucleus.
The first quantum-mechanical treatment of the Rutherford model was given in 1913 by Bohr.
According to the Bohr theory the energies and radii of the orbits are (in Gaussian units)
\begin{equation}
 E_{s} = - \frac{\mu e^4}{2 \hbar^2 s^2 }~; \qquad a_s = s^2 a_B; \qquad a_B =
\frac{\hbar^2}{\mu e^2}; \qquad \mu = \frac{m_e m_+}{m_e + m_+}.
\end{equation}
Here and in the following we denote the so-called main quantum number by $s$ in order not to be mixed with the density $n$.
According to the Bohr model there are infinitely many levels close to the series limit $s \rightarrow \infty$.
A first statistical theory of the Bohr model was developed already in 1913 by the Austrian physicist Karl Herzfeld \cite{Herzfeld}. Herzfeld detected that the Bohr model had a serious deficiency, the internal atomic partition function
\begin{eqnarray}
\sigma(T) = \sum_{s,l,m}\exp\left( -\beta E_s \right) = \sum_{s=1}^{s_m} s^2 \exp\left(\frac{I}{s^2 k_B T}\right); \qquad I = \frac{\mu e^4}{2 \hbar^2 s^2 } ~.
\label{eq:PlasmenZustandssumme1}
\end{eqnarray}
is divergent ($I=-E_1$: ionization energy).

\begin{figure}[htbp]
\includegraphics[width=4cm,height=5cm,angle=0]{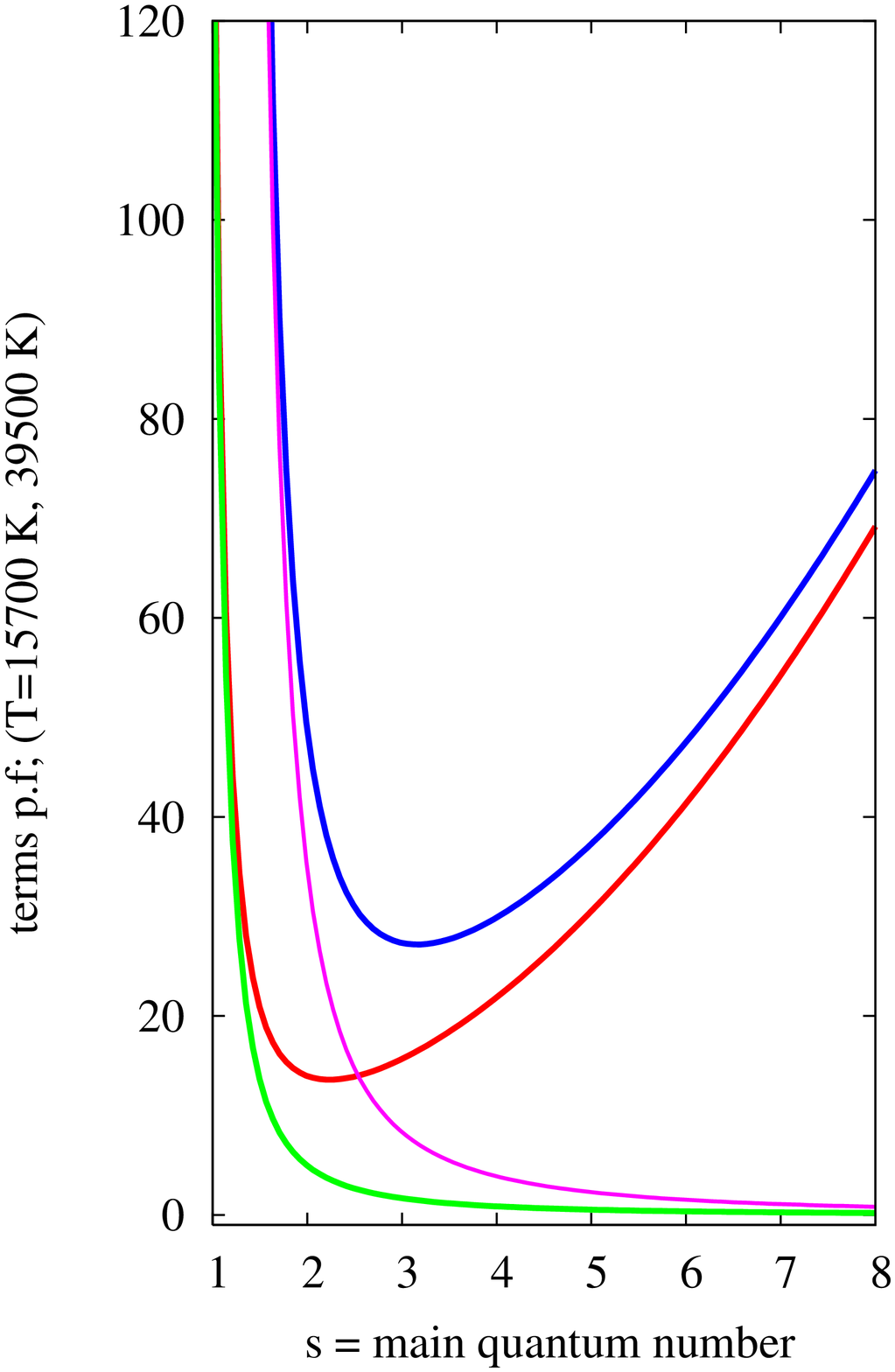}
\includegraphics[width=6cm,height=5cm,angle=0]{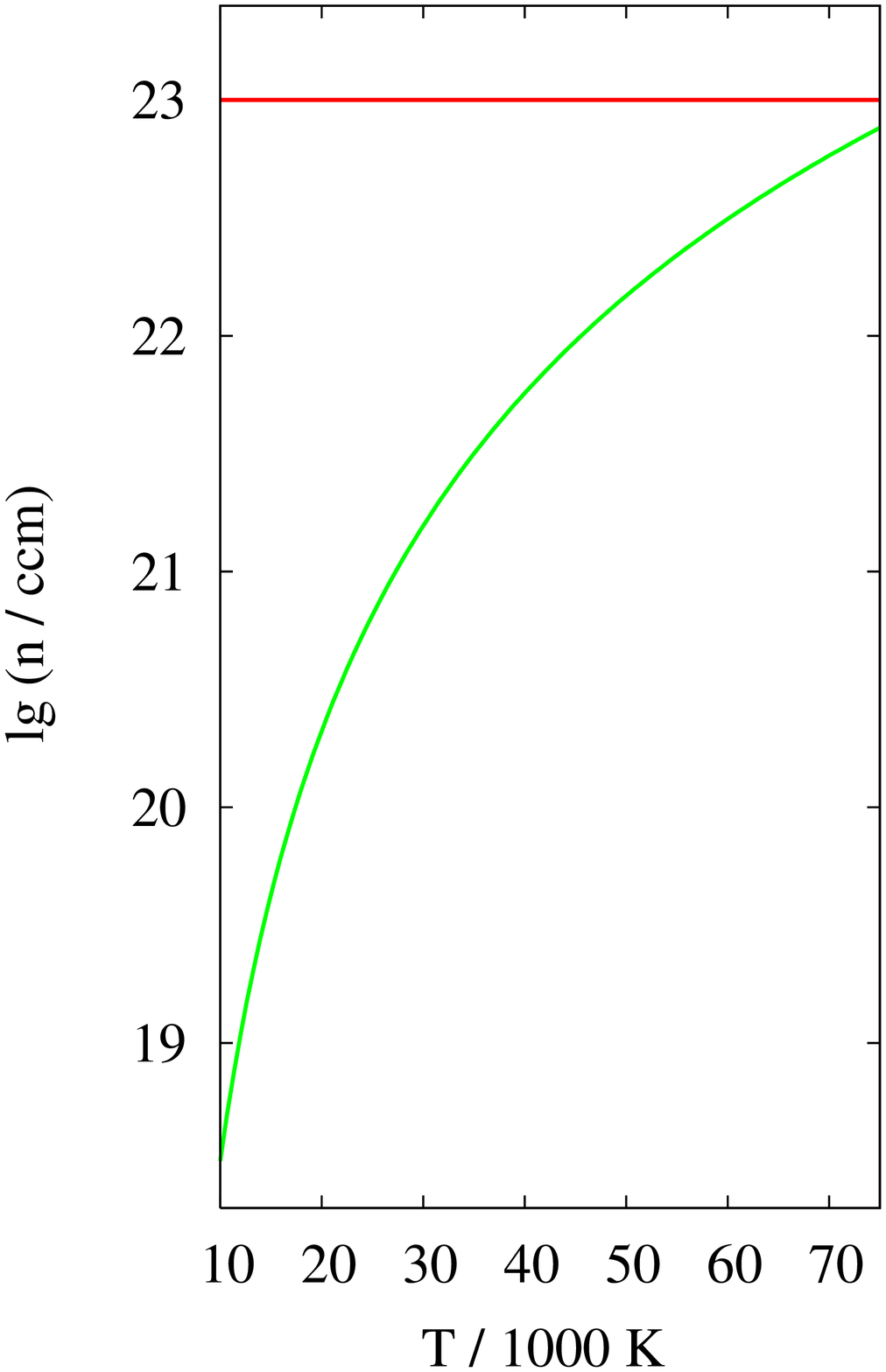}
\caption{Left panel: The terms in the atomic partition function defined by Eq. (\ref{eq:PlasmenZustandssumme1})
for two temperatures corresponding to $\beta I = 5$ (red and green curves above) and $\beta I = 10$
(blue and purple curves below).
Up to terms of order $E_{s_m} \simeq k_B T$ the terms in the usual partition function decrease monotonically and then they start to increase as $s^2$ leading to the Herzfeld divergence problem. For comparison we show the Brillouin-Planck-Larkin renormalized partition function with the terms of a convergent series. Right panel: The region in the density-temperature plane of Hydrogen where the formation of bound states is expected, see text.
}
\label{Zustandssumme}
\end{figure}

We see that the contributions to the internal atomic partition function for Hydrogen according to the definition by Eq. (\ref{eq:PlasmenZustandssumme1}) increase as $s^2$ (see Fig. \ref{Zustandssumme})
and the sum can get any value in dependence on the maximal number $s_m$ and is
divergent. This problem exists for all elements.
It is inherent in any Coulomb system, e.g. the electron - hole plasma in excited
semiconductors that will not be considered here.
Here we deal mainly with Hydrogen which is the most abundant element in the
Universe, its  physical properties raised, in the
past, many works.

The development of the quantum statistics of plasmas is intimately connected with the requirements
of a new scientific discipline - astrophysics. In order to bring some systematics into the existing
observations, to understand why spectra appeared under definite conditions, a theory was required which
provides the abundancies of atomic energy levels in stars. Such a theory was created on the basis of the
Rutherford-Bohr model by the work of Eggert and Saha around 1920. Their work is an application of thermodynamics
to the Rutherford-Bohr model, which was inspired by the papers of the Nernst school. The ionization problem was studied within the
chemical picture and using Sackur's and Tetrode's statistical expression for the entropy
of gases. Eggert published a first approach in 1919 \cite{Eggert}, his paper
came to the attention of the young bengalic physicist Saha, who succeeded to
improve the approach considerably \cite{Saha20,Saha21}.
Saha considers, in the paper from 1921,
Calcium plasmas as an example. Formulating his equation for Hydrogen we get for
the fraction of ionized atoms $\alpha$ the relations
\begin{equation}
\frac{1-\alpha}{\alpha^2} = n K ;\qquad K = \Lambda^3 \exp(\beta I); \qquad \Lambda = \frac{h}{\sqrt{2\pi \mu k_B T }}.
\end{equation}
The solution of this quadratic equations is given by
\begin{eqnarray}
\alpha = A(\gamma), \qquad \gamma= n K(T), \qquad A(x) = \frac{1}{2 x}\left[\sqrt{1+4 x} - 1 \right].
\label{alpha}
\end{eqnarray}
Considering an ideal mixture of electrons, ions and atoms we obtain for the pressure
\begin{equation}
\beta p = (1 + A(\gamma)) n.
\end{equation}
We call the function $A(\gamma)$, which monotonically decreases with the
density, the Saha function. \\
At high densities where the mean distance of protons approaches the Bohr radius,
this leads to an enormous pressure acting on the neutrals which will finally
be destroyed; we note however that this limit is not described in the present approach.
The region were bound states may exist is shown in Fig. \ref{Zustandssumme}.

Eggert and Saha used the thermodynamic approach of Sackur and Tetrode which needs only the binding energy, avoiding
this way the problem of divergence of the partition function.
For more accurate treatment of astrophysical tasks the most important problem
was the divergence of the partition function of the Bohr atom.
As already said above, the terms in the atomic partition function for Hydrogen according to the
definition by Eq. (\ref{eq:PlasmenZustandssumme1}) diverge as $s^2$ (see Fig.
\ref{Zustandssumme}).
The easiest way is to cut the sum at the lowest terms corresponding to $|E_s|
\simeq k_B T$. This way of solving the divergence problem is due to the
classical paper of Planck published in 1924 \cite{Planck}, later in 1938 applied
to special plasma problems by Riewe and Rompe \cite{RieweRompe}. Developing
Plancks approach Brillouin proposed in 1932 a more smooth
way of removing the divergence which led to the formula
\begin{eqnarray}
\sigma_{\rm BPL}(T) =  \sum_{s=1}^{\infty} s^2 [\exp(\beta I/s^2) - 1 -
(\beta
I / s^2)].
\label{BPLZustandssumme}
\end{eqnarray}
This procedure was justified only in 1960 by Larkin \cite{Larkin} and subsequently by the present
authors in collaboration with D. Kremp by using strict quantum-statistical methods
\cite{EbAnn67,EbPhysica,KrKr68,EbKrKrCPP70,RedBook,GreenBook}. For
later purpose it is interesting to note, that this so-called
Brillouin-Planck-Larkin (BPL) partition function
may be expressed (just after developing the exponential function) by
a different series which contains the Riemann $\zeta$ function
\begin{eqnarray}
\sigma (T) = \sum_{k=2}^\infty \frac{\zeta(2k -2) (\beta I)^k}{k!}.
\label{BPLexp}
\end{eqnarray}
In the present paper we will give a systematic quantum-statistical statistical theory of matter within
the Rutherford model, including in particular the equation of state.
We concentrate on the so-called physical picture which stays within the Rutherford picture and avoids the introduction of chemical species like atoms, molecules etc.
We note that in some earlier work the Saha theory has been
proven to become asymptotically exact in zero-temperature {\it and}
zero-density limit \cite{Fefferman,Macris,AlastueyBaCoMa03}.
Here we restrict ourselves mainly to a region of practical interest,
i.e., to temperatures between $20\, 000$ K and $100\, 000$ K
and particle densities between $10^{18}$ and $10^{22}$ per cm$^3$.
In some of our earlier work the transition to a chemical picture was made following the principle of equivalence that bound states are to be treated
on the same footing as free particles \cite{EbPhysica,RedBook,EbKrKrRo86}. This approach was quite successful in the description of partial ionization
\cite{RedBook,EbRi85,Chabrier,befjnrr,befjrr99,EbHaJRR5}. On the other hand, one has to accept that in the region of
high densities several serious difficulties appear which are mainly connected
with the impossibility of a clear distinction between free and bound states
in a dense system \cite{EbHi02}.
This is the reason why we stay here within the physical picture taking into account the important
higher order terms \cite{BobrovTrig10}.
In a similar spirit, but using different techniques, a consistent treatment of bound state contributions within the physical picture
has been constructed already by Alastuey et al. \cite{AlastueyPe,AlastueyBaCoMa08,Alastuey}.
We show here first a treatment of the Coulomb singularities in the framework of the virial expansions within the canonical ensemble, then we discuss the fugacity expansions and identify the higher order contributions
which allow a Saha-type approximation. In the last part we investigate the role of the identity of electrons within the grand canonical ensemble.

In the first part of this work we will to show that the Brillouin-Planck-Larkin procedure provides the most natural
way to avoid the divergencies for non-degenerate systems. In the last part we study degenerate systems and show that a consequent treatment of the identity of electrons in the Hartree-Fock approximation including bound states is a key for the treatment of
hydrogenic bound states. The present approach is based on the method of Green's
functions as developed in
\cite{RedBook,GreenBook,EbKrKrRo86,RedmerReport,MottBook}.

Any successful description of plasmas has to go beyond the Saha theory which is something like the lowest density order approximation
for plasmas. This way our strategy should be to obtain first an approximation
equivalent to the Saha description and then to go further. The derivation of
Saha-type approximations from quantum statistics was first studied by Planck,
Fowler and Brillouin \cite{Planck,Fowler,Brillouin}. Since then it is a topic of
permanent interest due to the importance of Saha-type equations for experimental
work
and for many technological applications. A first big progress in the strict
quantum statistical treatment was achieved in the 60th by papers of Montroll,
Ward, Vedenov, Larkin, Abrikosov and others \cite{Larkin,Vedenov,Abrikosov} and
in subsequent work
\cite{EbAnn67,EbPhysica,EbKrKrCPP70,RedBook,GreenBook,
EbKrKrRo86,KrKrEbPhy71,KSKBook,
wittsaka,riewitt}.
We note that a consistent description of bound states
in macroscopic physical systems is a hot problem of modern quantum statistics and was
treated also independently and using different techniques by other workers \cite{AlastueyPe,AlastueyBaCoMa08,Alastuey}.

\section{The model of Planck for Hydrogen plasmas}

Planck's work was inspired by the earlier papers of Eggert and Saha that was an application of thermodynamics and Sackur's and Tetrode's expression for the entropy. Accordingly Eggert and Saha needed only the binding energy, avoiding
this way the problem of divergence of the partition function.
In his Annalen paper from 1924, Max Planck uses a more consequent statistical approach based on the partition function.
He splits first the atomic partition function into three contributions
\begin{equation}
\sigma(T) = \sigma_3(T) + \sigma_2(T) + \sigma_1(T).
\end{equation}
where $\sigma_3(T)$ is the contribution of the lowest bound states $|E_s| > k_BT$, $\sigma_2(T)$ is the contribution of the
bound states below the series limit and $\sigma_1(T)$ is the entire remainder.
Using a rather complicated quasiclassical derivation Planck was able to show
that due to compensation effects it holds
\begin{equation}
\sigma_2(T) + \sigma_1(T) \simeq \frac{V}{\Lambda^3}.
\end{equation}
In other words the sum of the second and the third contributions reduces to the partition function of an ideal gas,
i.e. the all diverging contributions cancel each other. Planck's paper was
considered in those years to be very important, so it was
discussed in some detail in several textbooks and monographs, e.g. in the books written by  Brillouin \cite{Brillouin}
and Fowler \cite{Fowler}. We follow here mostly Brillouin. According to Brillouin \cite{Brillouin}, the cancellation effect found by Planck is exactly true only if one makes the particular choice
\begin{equation}
\sigma_3 (T) = \sigma_{\rm BPL}(T).
\end{equation}
As the derivations of Planck, Fowler and Brillouin are very
difficult to understand we
give the proof of Planck's important statement in a more easy version which however uses the same physical assumptions
\cite{EbAnn67}. We start with Planck's final expression
\begin{eqnarray}
F = k_B T N [\alpha [\ln (n \Lambda_e^3 \alpha)+1]+ \alpha [\ln (n \Lambda_i^3
\alpha)+1]\nonumber\\
+ (1-\alpha) [\ln (n \Lambda_a^3 (1 - \alpha)) - \ln (\sigma_3(T)) + 1] ] +
F_{\rm ex}.
\end{eqnarray}
Here $F_{\rm ex}$ is the excess part of the free energy. Planck and Brillouin claim that this contribution is zero, $F_{\rm ex}=0$; however these researchers
did not include screening, an effect detected nearly at the same time by Debye. In the following part of this section we repeat the derivation
of the excess part but include also the contributions from screening effects using
a semiclassical screened cluster expansion approach \cite{EbAnn67,Friedman}.
In order to find the excess free energy of a plasma we have to  calculate
all contributions
beyond the the ideal contributions of electrons, protons and atoms. Following
\cite{EbAnn67,EbPhysica} and
using the classical technique of screened virial expansions \cite{Friedman} we find including the Debye limiting law and the second
virial coefficents
\begin{eqnarray}
F_{\rm ex} = - k_B T V \left[\frac{\kappa^3}{12 \pi} + n_e^2 {\tilde B}_{ee} + 2
n_e n_i {\tilde B}_{ei} + n_i^2 {\tilde B}_{ii} \right] + \dots, \\
{\tilde B}_{ab} = 2 \pi \int dr\, r^2 \left[S_{ab}\exp(g_{ab} + \beta V_{ab}) - 1 -
g_{ab} - \frac{1}{2} g_{ab}^2\right].
\end{eqnarray}
The Boltzmann factors  without bound state contributions are given by
\cite{EbAnn67}
\begin{eqnarray}
S_{ee}(r) = S_{ii} = \exp(-\beta e^2 /r), \qquad V_{ab} = e_a e_b / r, \qquad g_{ab} = V_{ab} \exp (-\kappa r) \\
S_{ie}(r) = \exp(\frac{\beta e^2}{r}) - \frac{4 \pi}{\sqrt{2 \pi \mu k_B T}} \int_0^{p_0} dp\, p^2 \left[\exp(-\beta E(p,r)) -1 + \beta E(p,r) \right],\\
E(p,r) = \frac{p^2}{2 \mu} - \frac{e^2}{r}; \qquad p_0 = \sqrt{\frac{2 \mu
e^2}{r}}; \qquad \kappa^2 = 8 \pi n \beta e^2.
\end{eqnarray}
Here the second contribution in $S_{ie}$ is just subtracting the bound state  contribution from the full
Boltzmann factor in a classical approximation. Up to the Debye contribution, which was not yet known to Planck, since is was just
in print when Planck wrote his paper, this is just what Planck and Brillouin calculated.
After tedious calculations one can show analytically with a numerical check \cite{EbAnn67}, that the sum of the integrals
beyond the Debye term cancel each other and gives zero.
\begin{eqnarray}
F_{\rm ex} = - k_BTV \left[\frac{\kappa^3}{12 \pi} + {\cal O}(n^{5/2}) \right].
\end{eqnarray}
The excess free energy consists in the given quasiclassical approximation only of the Debye term and
some higher order corrections. In particular, there is no contribution in the quadratic order.
This way we could justify the remarkable result of Planck and Brillouin, we just found a correction, the Debye term.
Including this essentially classical term we get
\begin{equation}
\beta p = (1 + \alpha) n  -  \frac{1}{2}(8 \pi \beta e^2)^{3/2} (\alpha
n)^{3/2},
\end{equation}
$\alpha=A(\gamma)$ according Eq. (\ref{alpha}). A typical picture of the pressure in the Saha approach shows that bound states exist only
in certain valley in the density-temperature plane, see Fig. \ref{Zustandssumme}.
We will see later that the bound state contributions as well as the Debye contributions
are responsible for a reduction of the pressure in comparison to the ideal pressure (see Fig. \ref{presse}).
The Saha equation and the corresponding pressure do not describe the region of high densities, the Debye term yields just a first
correction. At higher densities another important influence is due to the Fermi pressure in the region of
electron degeneracy. In order to check for the influence of electron degeneracy
we compare with the ideal gas pressure including the Fermi pressure of the electrons:
\begin{equation}
\noindent \beta p_e^{F} (\beta, n) =  n[1 + 0.08839 n \Lambda_e^3 - 0.00083 n^2 \Lambda_e^6
+ 0.000012 n^3 \Lambda_e^9 + \cdots]; \,\, \Lambda_e =
\frac{h}{\sqrt{2 \pi m_e k_BT}}.
\label{idepress}
\end{equation}
Looking at typical isotherms (see e.g. Figs. 2, 3) we see that the pressure related to the ideal pressure
decreases with growing density and the pressure increases again. This
demonstrates that the region of bound state formation where the pressure is
significantly smaller than the ideal pressure of an electron-proton gas is a
limited region in the density-temperature plane (the valley of bound states).

The main task of this work is the study of the two transitions:\\
(i) the formation of bound states occurring at the cross-over from fully ionization to the region
were bound states are formed (we have partial ionization) observed with increasing density,\\
(ii) the destruction of bound states beyond the valley of bound states in the high density region.

The region of our interest in the density-temperature plane where bound states appear is demonstrated
in Fig. \ref{Zustandssumme}. We determine here the first transition by the densities where
$p / 2 n k_BT$  decreases below the value $3/4$ and identify the second transition by the condition that
$p^F / 2n k_B T$ exceeds the value $1$,
that means the Fermi pressure $p^F$ gives the overall dominant contribution.
From the physical point of view we should expect that the real pressure should be
nearer to the highest contribution, i.e at lower densities nearer to the pressure given by the Saha equation
and at high densities nearer to the Fermi pressure.
Later we will confirm the result that bound states occur only on an island in the $T-n$ plane.
In our primitive extended Saha model, the formation of atoms is described by the mass action law
which is part of the Saha model and the destruction at high density is described
in a very rough approximation only by the strong increase of the ideal Fermi
pressure. In the following sections we will show how these effects are more correctly described
by methods of quantum statistics. We present in Fig. \ref{Zustandssumme} first a horizontal line
corresponding to the density of $n = 10^{23}$ cm$^{-3}$.
For such a density the the mean distance between protons (or electrons) defined by
\begin{equation}
\frac{4 \pi}{3} n r_0^3 = 1
\end{equation}
is around the Bohr diameter.
\begin{equation}
r_0 \simeq 2 a_B; \qquad n a_B^3 \simeq 0.03.
\end{equation}
For larger densities (smaller average distances) the formation of bound states may be excluded,
the plasma behaves nearly as an ideal Fermi gas since there is not enough space for forming
atomic orbitals or molecular orbitals.
The disappearance of bound states is referred to as Mott condition,
corresponding to the fact that
atoms are destroyed if the mean distance of the electrons is much smaller than the Bohr radius.
Such situation was discussed in \cite{MottBook,EbBlRRR09,EbKrRo11}. For more
detailed information see \cite{GreenBook,KSKBook}.  We show, in Fig.
\ref{Zustandssumme}, a line $\alpha \simeq 0.5$. The Saha equation provides for
the condition $\alpha = 0.5$ the estimate
\begin{equation}
\gamma_0 = n K(T) = n \Lambda^3 \exp(I/k_B T) = 2.
\end{equation}
We are here mainly interested in the region of atomic bound states.
Near to the upper border of the corridor region shown in Fig. \ref{Zustandssumme} only the atomic ground state exists,
but near to the lower border many bound states are to be expected and we have to avoid the divergence
of the partition function by appropriate renormalization.
At lower temperatures $T \le 30\, 000$ K we may expect also the formation of molecules in a certain density region.

\section{Derivation of the Brillouin-Planck-Larkin partition function by low-density virial expansions}

Following our earlier papers \cite{EbKrKrCPP70,KrKrEbPhy71} and books \cite{RedBook,GreenBook}
we use the method of cluster expansions in order to derive the exact virial functions.
We start here with a treatment of the Coulomb contributions in the framework of the virial expansions within the canonical ensemble.
For a real quantum gas with only short range forces the free energy
may be at low density described by a virial expansion \cite{EbKrKrCPP70,KrKrEbPhy71}
\begin{equation}
F = F_{\rm id}^B - k_B T V \left[\sum_{ab} n_a n_b B_{ab} + \sum_{abc} n_a n_b n_c
B_{abc} + \cdots \right].
\end{equation}
\begin{equation}
B_{ab} =   {\rm const}\,\,  {\rm Tr} [\exp (-\beta H_{ab})  - \exp (-\beta H_{ab}^0)] ,
\end{equation}
where the Hamiltonian of pairs of species $a$ and $b$ is defined as
\begin{equation}
H_{ab} = - \frac{\hbar^2}{2 \mu_{ab}} \Delta + V_{ab}.
\end{equation}
Using the resolvent representation for the exponential operator we get
\begin{eqnarray}
B_{ab} &=&\frac{4 \pi^{3/2}(1+\delta_{ab})
\lambda_{ab}^3}{(2s_a+1)(2s_b+1)}\frac{1}{2\pi \,i}\int_c \exp(-\beta z) F(z) dz,
\qquad \lambda_{ab} = \frac{\hbar}{\sqrt{2 \mu_{ab} k_B T}},\nonumber\\
F(z)& =& {\rm Tr} \left[\frac{1}{H_{ab} - z } -\frac{1}{H_{ab}^0 - z
}\right].
\end{eqnarray}
After some transformation we arrive at the following representation by Jost
functions $D_{\ell}(z)$
\begin{eqnarray}
F(z)=\frac{(2s_a+1)(2s_b+1)}{(1+\delta_{ab})}\sum_{\ell=0}^{\infty}(2\ell+1)
\left[1\pm\delta_{ab}\frac{(-1)^{\ell}}{(2s_a+1)}\right](-1) \frac{d}{dz}\ln
D_{\ell}(z).
\end{eqnarray}
The Jost functions $D_{\ell}(z)$ are analytical functions with poles at the
bound
states and a branch cut at the positive real axis  defined by the scattering
phase shifts. These functions as well as other scattering quantities are exactly known for Coulomb systems.
However in the procedure of calculating the integrals there  is  a difficulty
due to the long range character
of Coulomb forces and the necessity to introduce some screening procedure.
In analogy to the classical case discussed in the previous section we may write including screening effects
\begin{equation}
F = F_{\rm id}^B - k_B T V \left[\frac{\kappa^3}{12 \pi} + \sum_{ab} n_a n_b {\tilde
B}_{ab} + \sum_{abc} n_a n_b n_c {\tilde B}_{abc} + \cdots \right].
\end{equation}
Here the tilde denotes that screened potentials were included into the virial coefficients.
In order to reduce the difficulties connected with a complete derivation
  we proceed as follows: The screening procedure is necessary only for the
lowest orders of the interaction parameter $e^2$. Higher orders in $e^2$
beginning with $e^8$ have already a decay rate of $ e^8 / r^4$
and do not need any screening. Taking into account that screening refers to the low orders we split the second virial coefficients into two parts
\begin{equation}
{\tilde B}_{ab} = G' {\tilde B}_{ab}   + G'' {\tilde B}_{ab}; \qquad G'' {\tilde
B}_{ab} = G'' B_{ab},
\end{equation}
\begin{eqnarray}
G' {\tilde B}_{ab} = O(1) + O(e^2) + O(e^4) + O(e^6), \qquad G'' B_{ab} = O(e^8)
+ O(e^{10}) + \cdots .
\end{eqnarray}
Let us first calculate the higher orders, remembering that according to Brillouin and Larkin, here the contribution of the bound states are to be expected. Introducing here the known Jost functions, the final result without symmetry effects reads
\begin{eqnarray}
G'' B_{ab} = 2 \pi^{3/2} \lambda_{ab}^3 [1 + \beta e_a e_b \kappa]  \sum_{m \ge
4} \frac{\xi_{ab}^m}{2^m \Gamma(m/2+1)}[\zeta(m-2)
\pm \delta_{ab} \frac{(1 - 2^{2-m})}{2 s_a +1} \zeta(m - 1)]\,.
\end{eqnarray}
The contributions in lower order have to be calculated term by term. The result
of tedious calculations is ($C=0.577$ -- Euler's constant)
\begin{eqnarray}
G'{\tilde B}_{ab}&=& 2 \pi \lambda_{ab}^3 [1 + \beta e_a e_b \kappa] [-
\frac{1}{6}\xi_{ab} - \frac{\sqrt{\pi}}{8} \xi_{ab}^2
- \frac{1}{6}(\frac{1}{2}C + \ln 3 - \frac{1}{2}) \xi_{ab}^3 ]\nonumber\\
&+& \frac{\pi}{3}(\beta e_a e_b)^3 [(1 + \beta e_a e_b \kappa) \ln (3 \kappa
\lambda_{ab} -
\beta e_a e_b \kappa (1 - \ln (4/3)]\nonumber\\
&&\pm \delta_{ab}[1 + \beta e_a e_b \kappa]\left[\frac{\sqrt{\pi}}{4} +
\frac{\xi_{ab}}{2} +  \frac{\sqrt{\pi} (\ln 2) }{4} \xi_{ab}^2 +
\frac{\pi^2}{72} \xi_{ab}^3 \right].
\end{eqnarray}
Note that most difficulties are connected here with the term of order $\xi_{ab}^3$ \cite{RedBook,GreenBook,KSKBook}.

For charge and mass symmetrical systems (all masses are equal and the reduced masses are just half of it)
the odd contributions cancel
and we get in the case of Hydrogen for the sum, taking into account $\beta I = \xi^2 / 4$:
\begin{eqnarray}
\sum_{ab} G'' B_{ab} = 2 \pi \sqrt{\pi} \sum_{k=2}^{\infty} \frac{\zeta(2k-2)
\xi^{2k}}{2^{2k} k! } = 4 \pi \sqrt{\pi} \lambda_{ie}^3 \sigma_{\rm BPL} (T).
\end{eqnarray}
This way we have given a strict derivation of the Brillouin-Planck-Larkin partition function given by Eq. (\ref{BPLZustandssumme}).
The final results for the equation of state of Hydrogen may be expressed in form of a
density expansion :
\begin{eqnarray}
\beta p (\beta, n) &=&  2 n  - 2 n (8 \pi \beta e^2)^{3/2} \sqrt{n}\nonumber\\
   &-& 2 n^2 [\Lambda^3 \sigma_{\rm BPL}(T)][1 + \frac{3}{2} \beta e^2 \kappa)]
- 2
n^2 K^*(T)] + L(T) n^{5/2} + {\cal O}(n^3 \log n).
\label{pressPBL}
\end{eqnarray}
The function $K^*(T)$ which appears in the formula Eq. (\ref{pressPBL})
depends only of the temperature. This transcendent function is
exactly known \cite{RedBook,GreenBook,KSKBook} and may be calculated as
an infinite series in $\xi$ using the expressions for $B_{ab}$ given above.
There are also tables for Hydrogen available \cite{RedBook}.
However since the full expression is a quite long and not easy to handle,
we will give here only a rather good (in numerical respect) and simple approximation
\begin{equation}
K^*(T) = - \frac{1}{2}\Lambda^3 \beta I - \frac{1}{8 \sqrt{2}} \Lambda^3.
\label{K(T)}
\end{equation}
This expression is exact with respect to the largest asymptotic term
${\cal O}(I/k_B T)$ and was checked
with the available tables \cite{RedBook}. As a technical remark we note, that the
terms not contained in the approximation eq.(\ref{K(T)}) are mainly given by the contributions of the odd powers in $\xi$ to the pressure.
We do not give here a detailed discussion since these terms give only a small
contribution to the pressure in the region $10^4 {\rm K} < T < 10^5$ K. The exact expression may be found in
original articles \cite{EbPhysica,EbKrKrCPP70} and in several books \cite{RedBook,GreenBook,KSKBook}.
The complete term of order $n^{5/2}$ has also been computed in several works \cite{wittsaka,Kahlbaum00,alasper}. We will not discuss this matter here since we use here some approximations for $L(T)$ and because
this function does not contain bound state contributions.
There were different attempts to determine such contributions $n^{5/2}$
in the framework of Green's function techniques. Here we mention \cite{wittsaka,riewitt}. These results should
be compared to the corresponding expressions for the $n^{5/2}$ terms given by \cite{alasper,brownyaffe}.
Though there are results which seem to be in agreement with each other, there remains still some
open questions \cite{EbKrRo11}, see \cite{kraeft05}.
Here we are interested mainly in the bound state effects and restrict ourselves to the
asymptotic results for the case $\xi_{ie}^2 \gg 1$ what is equivalent to $T \ll I / k_B$.
For the case of Hydrogen and lower temperatures $T \ll I / k_B$ we get in asymptotic approximation the final results for the pressure at small densities:
\begin{eqnarray}
\beta p (\beta, n) &=& n [1 + \frac{1}{8 \sqrt{2}} n \Lambda_e^3 + \cdots] + n
- 2 n
\frac{1}{2}(8 \pi \beta e^2)^{3/2} \sqrt{n}\nonumber\\
&-&   n^2 \Lambda^3 (\beta I) + n^2 \Lambda^3 \sigma_{\rm BPL} (T) [1 +
\frac{3}{2}\beta e^2 \kappa] +\cdots + {\cal O}(n^{5/2}) + {\cal O}(n^3 \log n).
\label{pressPBL0}
\end{eqnarray}
This way we have shown that in the quantum-statistical expressions for the thermodynamic functions appears
the Brillouin-Planck-Larkin partition function in a natural way as a convergent infinite sum starting with terms of order $e^8$.
As mentioned in the previous section, a first approximation to this partition function was obtained by Planck \cite{Planck} and a first explicit derivation was given by Brillouin \cite{Brillouin}. Later an asymptotic quantum-statistical derivation was given by Larkin \cite{Larkin}.
We follow here Larkin's asymptotic approach which was extended later in \cite{EbPhysica} but include more terms than in the earlier work.
Exact expressions for the orders ${\cal O}(n)$ and ${\cal O}(n^{5/2})$ in the pressure including all orders in $e^2$ were derived by using cluster expansion method and Green's function methods in
\cite{EbPhysica,EbKrKrCPP70,RedBook,GreenBook,KrKrEbPhy71,KSKBook} and
by using different techniques in \cite{AlastueyPe,AlastueyBaCoMa08,Alastuey}. We will not give a complete survey about the existing
exact results about the lower orders in the density but will instead direct the following study on the most important
higher order terms. However let us still make a remark about the foundation of the Brillouin-Planck-Larkin partition function.
There seems to be a never-ending discussion about the Brillouin-Planck-Larkin partition function.
Just to give one example, Starostin and Roerich \cite{Starostin}
derived recently a quite different expression.

Our point of view which is based in particular on the results obtained here is the following:\\
(i) The quantum-statistical result for the contribution ${\cal O}(n^2)$ to the pressure is exact. The result of several groups
working with independent methods give the same results (see e.g. \cite{GreenBook,AlastueyPe}).\\
(ii) The BPL-partition function is asymptotically in full agreement with the
second virial coefficient. There is a small freedom in the special choice of a
partition function, as far as the asymptotics is not violated. However so far no
good reason is to be seen, why
we should leave the BPL- partition functions and make another choice
\cite{EbHi02}.\\
(iii) According to Onsager's bookkeeping theorem, in a correct theory any small change of the partition function does not matter,
the final results for any physical observables are stationary with respect to
small changes \cite{EbHi02}.\\
Some of the critics of the BPL partition function is just based on intuitive arguments
and is not based on a serious revision of the quantum-statistical foundations given in the work of Vedenov and Larkin and the subsequent papers.

Admittedly the quantum statistics behind this formula is very complicated but nevertheless it is strict.
With respect to the intuitive arguments we only want to say, the formal reasons for having no terms of order $O(e^0)$ and ${\cal O}(e^4)$
in the Planck-Brillouin-Larkin formula is that these terms were
consumed already in the formulation of the ideal terms which are ${\cal O}(e^0)$ and the screening terms
${\cal O}(e^3)$, and ${\cal O}(e^4)$ which arise formally by summing up an infinite series in ${\cal O}(e^n)$. This procedure
consumes some  terms of order ${\cal O}(e^3)$, and all contributions of order ${\cal O}(e^4)$
to the partition function which cannot appear anymore in exact expression for the partition function, if double counting of terms
is strictly avoided. Another, formally equivalent argumentation is based on the
fact of compensation of the
terms below the series limit and above the series limit is most essential in this context \cite{KSKBook}.
We want to underline again that there is no unique choice of the partition function for Hydrogen, there are other possibilities \cite{EbHi02}. Essential is however, that all possible choices should be compatible with the
quantum statistical formulae given above. That means among the many admittable versions of partition functions,
the essential criterion for correctness is Onsager's bookkeeping rule. In other
words the question is: Is there a correct counting of all contributions or not.
Is there a check for double-counting or not \cite{EbHi02}. As a matter of fact,
all partition functions which work without a BPL- or equivalent subtraction of
first order terms may have the problem of double-counting of certain
diagrams of order ${\cal O}(e^n)$.

Using the asymptotic expression for $K^*(T)$, which is asymptotically exact for $k_B T \ll I$,  the pressure may be written in a different form
\begin{eqnarray}
\beta p (\beta, n) &=&  n [1 + \frac{1}{8 \sqrt{2}} n \Lambda_e^3 +\cdots]
+ n - \frac{\kappa^3}{24 \pi} [1 - \frac{3 \sqrt{\pi}}{8} (\kappa
\lambda)+\frac{3}{10}(\kappa \lambda)^2 + \cdots]\nonumber\\
&-& n^2 [\Lambda^3 \sigma_{\rm BPL}(T)][1 - \beta e^2 \kappa] + {\cal O}(n^{3}
\log n).
\label{pressPBL1}
\end{eqnarray}
We combined one of the two contributions from $K^*(T)$ with the ideal electron pressure and the
other one with the Debye law.
Further we introduced the mean thermal wave length
$
\lambda = \lambda_{ie}
$
which is an effective quantum length, in some correspondence to
the Debye-H\"uckel parameter in the classical theory
($\mu$ - reduced mass). We have to note here again that the results given above are not fully complete in the order $n^{5/2}$ and in missing
higher order corrections. However this problem seems to be not relevant for the bound state problem.
A formally equivalent form of Eq. (\ref{pressPBL1}) which shows better the general structure is the following
\begin{eqnarray}
\beta p (\beta, n) &=&  n[1 + \frac{1}{8 \sqrt{2}} n \Lambda^3+\cdots] + n - n
\gamma
[(1 - \beta e^2 \kappa +\cdots) + \cdots]\nonumber\\
&-& \frac{\kappa^3}{24 \pi} [1 - \frac{3}{2} \gamma +\cdots][1 - \frac{3
\sqrt{\pi}}{8} (\kappa \lambda)+ \frac{3}{10}(\kappa \lambda)^2 +\cdots] +
{\cal O}(n^{3/2}),
\label{pressPBL2}
\end{eqnarray}
\begin{eqnarray}
\label{gammaBPL}
\gamma = \Lambda^3 \sigma_{\rm BPL}(T) = 8 \pi^{3/2} \lambda_{ie}^3 \sigma_{\rm BPL}(T).
\end{eqnarray}

\begin{figure}
\centering{\epsfig{figure=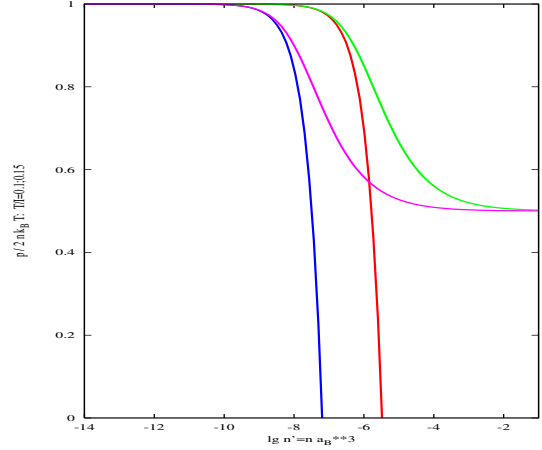,height=6cm,width=0.5\textwidth,angle=0}}
 \caption{ \label{presse}
Pressure related to the total classical pressure $2 n k_B T$ at constant temperatures $T = 0.1 I$ and $T=0.15 I$ in dependence on the density (log-scale). The result of density expansions up to 2nd order virial terms (red and green) are compared with corresponding fugacity expansions.
}
\end{figure}

Summarizing we may state that there exists an asymptotically exact expression for the pressure of Hydrogen in the limit of small densities, which is given in the formulae Eqs. (\ref{pressPBL1}), (\ref{pressPBL2}) which we have drawn in Fig. \ref{presse}. This approach is compatible with the extended Saha theory what encourages us to proceed in this direction. Let us first investigate Eq. (\ref{pressPBL2}) as it is, without completing the series.
Looking at the curves in Fig. \ref{presse} we see that the screening effects and the corresponding BPL effects tend to lower the pressure. We see that this lowering is too large at increasing densities.

The structure of Eq. (\ref{pressPBL2}) suggests that the expressions in the parenthesis may be
only the first terms of some infinite series with respect to $n \Lambda^3$, $\kappa$ and $\gamma$. Just to give a first example:
The first series is evidently nothing else than the ideal electron pressure, which may easily be extended to an infinite series
using the known expression for the ideal Fermi pressure.
There is no question that all these terms appear in the full cluster series.
In order to find the higher orders in the other series we have to use a more complicated procedure
which is based on fugacity expansions \cite{EbPhysica,RedBook,Bartsch}.
According to our general philosophy, the strong decay
of the pressure observed in second order density expansions (Fig. \ref{presse}) is due to the fact that
important higher order terms corresponding to some infinite series were omitted in the derivation based on the canonical ensemble.
To circumvent this difficulty is not an easy task. We mention that Alastuey et al. \cite{AlastueyBaCoMa08,Alastuey}
succeeded already in deriving a Saha type expression in summing up an infinite series of terms.
Here we follow a similar idea but go an alternative way through fugacity expansions. As shown much earlier, fugacity-like
expansions are equivalent to mass-action laws. This was used in plasma theory
by Bartsch, Ebeling and others \cite{RedBook,EbPhysica,Bartsch} and
worked out in large detail by
Rogers et al. \cite{Rogers}. We mention that Rogers et al. succeeded to show that extended fugacity expansions allow an excellent
description of measurements for the oscillation modes of the sun \cite{Dappen}.

\section{Combined density-fugacity expansions including nonlinear ring and ladder contributions}

At higher densities a systematic quantum statistical approach to the pressure of plasmas can
be given based on the Green's function representations of the pressure. The
central relation for the pressure reads \cite{RedBook,GreenBook,EbKrKrRo86}
\begin{eqnarray}
\label{qupress}
p(\beta, \mu_e,\mu_i)=p_{id} - \frac{1}{2 V} \int_0^{1} \frac{d \lambda}{\lambda} \int d 1 {d {\tilde 1}}
V(1 {\tilde 1}) G_2 (1, {\tilde 1},1^{++},1^+: {\tilde t}_1 = t_1^+),
\end{eqnarray}
where $1 = \{\vec p_1, \sigma_1 \}$ denotes momentum and spin variables.
The Green's function representation works in the grand canonical ensemble which provides us a series
in the fugacities
$$
z_a = \frac{2s_a + 1}{\Lambda_a^3}\exp(\beta \mu_a) = n_a \exp(\beta \mu_a^{\rm ex})
$$
instead of a series in the densities.
Note that $\mu_a^{\rm ex}$ is the excess part of the chemical potential (the part beyond the ideal Boltzmann term)
and that therefore $z_a \rightarrow n_a$ for small densities. At small fugacities (densities) the
pressure is given as an expansion in the fugacities
\begin{eqnarray}
\beta p = \beta \sum_a z_a + \sum_{ab} z_a z_b b_{ab}(\kappa_g) + \sum_{abc}
z_a z_b z_c b_{abc}(\kappa_g) +
\sum_{abcd} z_a z_b z_c z_d b_{abcd}(\kappa_g) + \cdots  ,
\end{eqnarray}
where
$$
\kappa_g^2 = 4 \pi \beta \sum_{a} z_a e_a^2
$$
is the grand-canonical screening length.
The structure of this series is similar to the density series, we have e.g. similar screening effects,
however there are some differences. e.g. the fugacity series contains more diagrams.
The contribution of bound states of an electron and an ion (atom)  is contained
in the contribution
$z_e z_i b_{ei}$, the contribution of molecules is contained in the term $z_e^2 z_i^2 b_{eiei}$.

General expressions for screened fugacity series were written down first by Montroll and Ward,
Vedenov and Larkin and explicit calculations were done by DeWitt and Larkin
\cite{Larkin,Vedenov,DeWitt}.
Semiclassical fugacity series were given by several workers \cite{RedBook,Bartsch}.

\begin{figure}
\centering{\epsfig{figure=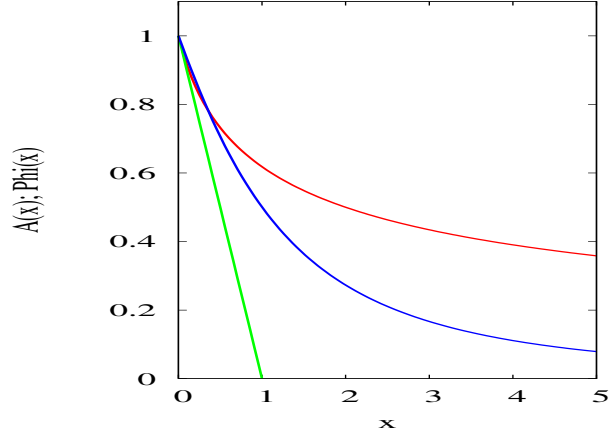,height=6cm,width=0.6\textwidth,angle=0}}
\caption{ \label{nonlfunc}
The nonlinear ring function $\phi(x)$, which determines the contribution of screening to the pressure  (blue line). Further we show the nonlinear
ladder function $A(x)$ (red line) which determines the contribution of the bound states to the pressure
as well as the linear approximation, which gives only the lowest order to the second virial coefficient (thin line).
}
\end{figure}

We are interested here mostly in the bound state effect and may simplify our
calculations by using approximations allowed for the region $T \le 157\, 000$ K
where the bound states appear. Further in order to simplify the quite complicate
evaluation of these diagrams we restrict the theory to the
region of non-degenerate plasmas. As to be seen from the above formulae, the
difference of the masses of electrons and ions (protons) do not play a role in
the region $|\xi_{ab}| \ll 1$ and the plasma behaves asymptotically like a
system with equal masses and equal relative De Broglie wave lengths
$
\lambda = \lambda_{ie}.
$
In the lowest approximation corresponding to Eq. (\ref{pressPBL1}) we find for
low temperature Hydrogen plasmas the following results valid up to order $O(z^{5/2})$
\cite{RedBook,GreenBook}.
\begin{eqnarray}
\beta p = z_e +  z_i + \frac{\kappa_g^3}{12 \pi} f(\kappa_g \lambda)  + 8 \pi
z_ez_i \lambda^3 \exp[1 + \beta e^2 \kappa_g] \sigma_{\rm BPL}(T) + \cdots ,
\label{PressFug0}
\end{eqnarray}
where the grand-canonical screening quantity is now defined by
$$ \kappa_g^2 = 4 \pi \beta (z_e + z_i) e^2, $$ and the quantum-statistical ring
function is a transcendent function \cite{EbPhysica,RedBook}
with the series
\begin{equation}
f(x) = \left[1 - \frac{3 \sqrt{\pi}}{16} x  + \frac{1}{10} x^2 + \cdots \right].
\end{equation}
The relations between densities and fugacities
are given by
\begin{eqnarray}
n_i = z_i \frac{\partial \beta p}{\partial z_i} = z_i + \frac{\kappa_g^2}{16} \frac{\partial}{\partial \lambda} (\kappa_g \lambda f(\kappa_g \lambda)) +
4 \pi z_e z_i [1 + \beta e^2 \kappa_g +\cdots] \lambda^3 \sigma_{\rm BPL}
(T)\cdots,
\label{DensFugi}
\end{eqnarray}
\begin{eqnarray}
n_e = z_e \frac{\partial \beta p}{\partial z_e} = z_e + \frac{\kappa_g^2}{16}
\frac{\partial}{\partial \lambda} (\kappa_g \lambda f(\kappa_g \lambda)) + 4 \pi
z_e z_i  [1 + \beta e^2 \kappa_g + \cdots] \lambda^3 \sigma_{\rm BPL}(T)\cdots.
\label{DensFuge}
\end{eqnarray}
We will show that going by iterations from the fugacity variable to the density
this result
occurs to be equivalent to eq.(\ref{pressPBL1}). In order to proceed with this complicated system of equations we go to
the special case of non-degenerate hydrogen at lower temperatures $T \le I / k_B$ and satisfying
$n \Lambda^3 \le 1$. As shown above, in this region the differences
of the electron ion masses does not matter and we my assume $z_e = z_i = z$.
In this approximation, Eqs. (\ref{PressFug0}), (\ref{DensFugi}),
(\ref{DensFuge}) read
\begin{eqnarray}
\beta p = 2 z + \frac{\kappa_g^3}{12 \pi} f(\kappa \lambda) + 8 \pi z^2
\lambda^3 \sigma_{\rm BPL} (T) +\cdots \,, \\
n = z + \frac{\kappa_g^2}{16 \pi} \frac{\partial}{\partial \lambda} (\kappa_g
\lambda f(\kappa_g \lambda)) + 8 \pi z^2 \lambda^3
\sigma_{\rm BPL}(T)\cdots .
\label{pn}
\end{eqnarray}
In order to represent the pressure by densities, we neglect, in a first step,
the bound state contributions.
Expressing first the fugacities by densities we find
\begin{eqnarray}
z = n \exp[ - \frac{\beta e^2 \kappa}{2} G(\kappa \lambda)],
\label{Fugdens0}
\end{eqnarray}
where \cite{EbPhysica,RedBook}
\begin{eqnarray}
G(x) = 1 - \frac{\sqrt{\pi}}{4} x + \frac{1}{6} x^2 - \cdots.
\label{Fugdens00}
\end{eqnarray}
Then excluding the fugacities from the series step by step we arrive
at the representation
\begin{eqnarray}
\beta p = 2 n - \frac{\kappa^3}{24 \pi} \phi(\kappa \lambda).
\label{Pressdens0}
\end{eqnarray}
Here $\phi(x)$ is another transcendent function, related to $f(x)$ and $f'(x)$, the so-called pressure ring function which is an analogue of the Debye-H\"uckel ring function \cite{RedBook,Vedenov,DeWitt}. The first terms of the series expansion read
\begin{equation}
\phi(x)=[1 -  \frac{3 \sqrt{\pi}}{8}x + \frac{3}{10} x^2 - \cdots].
\end{equation}
A full representation of this quite complicated transcendent function may be obtained by methods similar as described in \cite{RedBook}.
We found the infinite series
\begin{equation}
\phi(x) = 1 - \frac{\sqrt{\pi}}{3} \sum_{k=1}^{\infty} \frac{(k+1)(k+3)
x^{2k-1}}{2^k k!} + \frac{1}{3}\sum_{k=1}^{\infty} \frac{(k+1)(k+3) x^{2k}}{2^k
(2k-1)!!}.
\end{equation}
The asymptotic behavior of this function is given by
\begin{equation}
\phi(x)= \frac{3}{x^2}.
\end{equation}
This suggests the convenient Pad{\'e} representations, which will be used in the
calculations
\begin{equation}
\phi(x)= \frac{24}{24 + 9 \sqrt{\pi} x + 8 x^2}; \qquad G(x) = \frac{1}{1 +
(\sqrt{\pi}/4) x}.
\end{equation}
A representation by error functions reads
\begin{equation}
\phi(x) = \frac{3 \sqrt{\pi}}{3 x} +\frac{4 \sqrt{\pi}}{x^3} - \frac{4 \sqrt{\pi}}{x^3} \exp[x^2 / 4] -
\frac{\sqrt{\pi}}{3 x} [1 - \Phi(x/2)] + {_1F_1}. 
\end{equation}
The saturating behavior of the nonlinear function $\Phi(x)$ in comparison to
its linear approximation is shown in Fig. \ref{nonlfunc}.

We summarize the results we obtained so far: By means of grand-canonical methods we succeeded to obtain
some infinite series in the screening parameter $\kappa$ instead of the first
linear and quadratic terms only, as we found in the density series,
(see Eq. (\ref{pressPBL2})). This summing up a series in $\kappa$ which leads to
a saturating function (see Fig. \ref{nonlfunc}) and improves very much the
behavior at larger densities as we will show.

In our representation Eq. (\ref{pressPBL2}), there appears also another series
expansion in the density parameter $\gamma$ .
In order to sum up this series in $\gamma$, we proceed in a similar way. In
order to understand the structure of the series
we study at first the bound state contributions, neglecting the contributions coming from the ring term and from the
terms due to degeneracy. Further we replaced the factor $1 - \beta e^2 \kappa$
by a more smooth exponential function $\exp(- \beta e^2 \kappa)$.
This yields the simple quadratic equation
\begin{eqnarray}
n = z + 8 \pi z^2 \lambda^3 \sigma_{\rm BPL}(T).
\end{eqnarray}
By the way, this formula shows already that the
fugacities of the electrons and the protons should be rather small in the bound state region where the partition function is large.
The quadratic equation for the fugacities can be easily solved.
This way we find, after elimination of the fugacity $z$ (alternatively using the
$A$ -- function):
\begin{eqnarray}
z^0 = n A(\gamma); \qquad A(x) = \frac{1}{2x}[\sqrt{1 + 4 x} - 1 ] ; \qquad
\gamma= 8 \pi n \lambda^3 \sigma_{\rm BPL}(T).
\end{eqnarray}
We note the following series expansion and large $x$ asymptotic of the nonlinear function $A(x)$:
\begin{eqnarray}
A(x) = 1 - x + 2 x^2 -\cdots; \qquad  A(x) = \frac{1}{\sqrt{x}} + \cdots.
\end{eqnarray}
The saturating behavior of the nonlinear function $A(x)$ in
comparison to its linear approximation is shown in Fig.\ref{nonlfunc}.
This shows that in the region of large partition functions the fugacities disappear as
\begin{eqnarray}
z^0 = \frac{1}{\sqrt{8 \pi n \lambda^3 \sigma_{\rm BPL}(T)}}.
\end{eqnarray}
This is important for the understanding, why the fugacity series has, in the
region of bound states, a better convergence than the density series.
The better convergence is due to the fact that the fugacities disappear in the region where the bound state contribution is large.\\
Introducing the zeroth step of iteration $z^0$ into the pressure we find
\begin{eqnarray}
\beta p^0 = z^0 + (z^0 +  8 \pi z^2 \lambda^3 \sigma_{\rm BPL}(T)) =
z^0 + n = n (1 + A(\gamma)).
\end{eqnarray}
This surprising representation tells us that the Saha function $A(x)$ appears in
the pressure in a
quite natural way through the fugacity expansion. We have introduced elements of the chemical picture without
using the notation of chemical species explicitely.
The new "semi-chemical representation" includes the nonlinear function $A(\gamma)$ which saturates
at large densities (see Fig. \ref{nonlfunc}).
The new representation contains all terms in the density up to $n^2$ as well as
several higher order terms in the density contained in the nonlinear
$A(\gamma)$  which is related to the Saha function defined by Eq.
(\ref{alpha}). We need this kind of functions
to reproduce an ideal Saha-type behavior.
The fugacity series contains more terms than the density series, so we may expect that some of the difficulties connected with density expansions, as the strong decrease of the pressure with increasing density shown in Fig. \ref{presse} may be avoided. Indeed,  a representation of the curve
corresponding to Eq. (\ref{PressFug0}) shows a more reasonable behavior with increasing density which
is much closer to the Saha-type behavior.
As the curves shown in Fig. \ref{presse} demonstrate, the pressure according to the fugacity expansion goes to saturation. This is due to the fact that the fugacity is not fixed, it decreases with increasing
value of the partition function and this way limits the growth. Evidently the fugacity expansions provide a more correct description of the bound state contributions as demonstrated in Fig. \ref{presse}.

We find that density as well as fugacity expansions have both advantages as well as disadvantages:\\
1) The density expansion describes well the screening effects but
it fails to cope with the diverging contributions from the screening terms and from the BPL-partition function.\\
2) The fugacity expansion corresponds to an infinite density series including
the partition function $\sigma$.
If $\sigma$ is large, then the fugacity goes to zero what guarantees even at large densities always
finite contributions to the pressure. This is true for the screening contributions and for
the bound state contributions; any strong increase of contributions
suppresses the fugacities. In general, the fugacity expansions are more
smooth.

In conclusion we may expect that the best representation is obtained by extended density expansions
which contain additional contributions corresponding to the important damping terms in the
fugacity expansions. We may expect that this procedure provides Saha-like terms.
This kind of mixed expansions combines the positive features of both expansions
avoiding the negative features. In order to demonstrate this we started here from the density expansions and used
the fugacity expansions mostly only for finding the right continuation of the infinite
series with respect to the $\gamma$- parameter and the $\kappa \lambda$ parameter.

\begin{figure}
\centering{\epsfig{figure=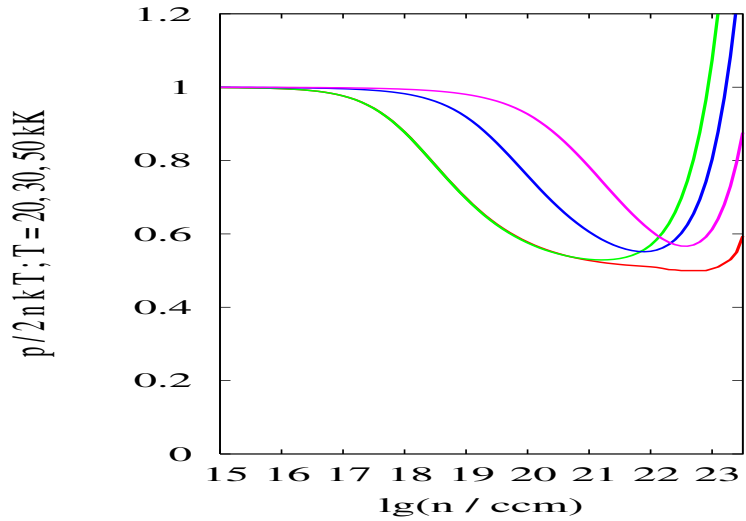,height=7cm,width=0.75 \textwidth,angle=0}}
\caption{ \label{qupress11}
Equation of state (EOS) calculated from an extended density representation including the nonlinear
ring and ladder functions $\phi(x),A(x)$ and including the Fermi pressure of the electrons.
We show the pressure related to the classical ideal pressure $2 n k_B T$
at constant temperatures $T = 20000, 30000, 50000$ K
in dependence on the density (in log scale). For the temperature $T = 20000$ K (lowest, green line) we give a comparison
with an earlier calculated curve based on the chemical picture and and Pad{\'e}
approximations \cite{EbRi85,befjnrr,befjrr99}.
}
\end{figure}

Following this line and including all terms up to known orders in the
densities and fugacities,
we obtain, in a first step, the following expression for the fugacities
\begin{eqnarray}
z = n a' ; \qquad a' = A \left( 8 \pi n \lambda^3 \exp[ - \beta e^2 \kappa
G(\kappa \lambda) ] \sigma_{\rm BPL}(T)\right).
\label{Fugdens1}
\end{eqnarray}
Note that here the second term plays the role of an effective density of the bound particles.
For the pressure we find, including again the full electronic Fermi pressure,
the following relatively simple formula which contains the three nonlinear functions
which we discussed in the foregoing part
\begin{eqnarray}
\beta p = p_e^{F} + n A(\Gamma) - \frac{\kappa'^{3/2}}{24 \pi} \phi(\kappa' \lambda),
\label{presscom}
\end{eqnarray}
with the definitions of renormalized values of the Saha parameter and the screening parameter
\begin{eqnarray}
\Gamma = 8 \pi n \lambda^3 \exp[ - \beta e^2 \kappa G(\kappa' \lambda) ]
\sigma_{\rm BPL}(T); \qquad \kappa' = \kappa A^{1/2}(\gamma).
\label{renomqu}
\end{eqnarray}
This provides us with a closed and relatively simple formula for the pressure, including the Fermi pressure.
The underlying assumption is that all electrons contribute to the Fermi pressure.
This result for the pressure according to the combined density-fugacity
representation Eq. (\ref{presscom}) which includes all contributions up to
second order in the fugacity is, for three temperatures, shown in Fig.
\ref{qupress11}.

The good convergence of the new expression including the nonlinear functions $A(x)$ and $\phi(x)$ based on the
fugacity series is explained by the fact that the fugacities of the electrons and the protons are rather small in the bound state region.
We notice again that this result comes from an extension of the density expansion including all quadratic
terms from the  fugacity virial expansions and the full Fermi pressure of the electrons. We see that the overall behavior of
the new representation is much better than that of pure density or pure fugacity representations. This way we may conclude that the most appropriate description of Coulomb systems is by density expansions and including some elements from the fugacity expansions. The physics behind is that the fugacity expansions describe well the saturable forces between bound states in Coulomb systems, but on the other hand the additive long range Coulomb forces and the screening effects are better represented by density series.

We note that the new theory based on Eqs. (\ref{presscom}), (\ref{renomqu}) is consistent with the Saha theory and in particular also with the  Saha-Debye-H\"uckel theory \cite{RedBook}. Comparing the curves based on the extended theory, Figs. \ref{qupress11} and  \ref{VglNewOld} with Fig. \ref{presse}, we see that the overall
behavior of the new extended density representations is much better than that of pure density
or pure fugacity representations.
Further we may state that the new extended analytical theory is, up to the
density $n \simeq 10^{21}$ cm$^{-3}$, in quite good agreement with
earlier calculations based an the chemical picture and quite complex numerical codes
\cite{EbRi85,befjnrr,befjrr99,EbHi02,befjrr01}

Note however that the agreement is based on a complete agreement in the lowest orders in the density, in higher orders several differences appear.
Beside the good agreement with the Saha theory and in particular with the extended
Saha-Debye-H\"uckel theory developed in \cite{RedBook} we list the following
deviations:\\
1. The exponent $\exp(\beta I)$ in the Saha equation is  replaced by the BPL-partition function in the quantum-statistical formula
 Eq. (\ref{presscom}).\\
2. The linear first Debye-H\"uckel screening function linear in $\kappa$ is
replaced here by a nonlinear ring
function $\phi(x)$ describing the quantum statistical ring sums in the grand canonical expansion in a good approximation.\\
3. The linear second virial coefficient is replaced by a nonlinear ladder function stemming from the representation
in the grand ensemble.

We note that the grand canonical ensemble plays an essential role in our
derivation, and further we note
that the formula Eq. (\ref{presscom}) corresponding to a second iteration is in some sense incomplete, e.g we would expect that in higher iterations
more $\kappa$-terms are replaced and that more terms corresponding to a mass-action law will appear.\\
We underline that several of the terms beyond $n^2$ and $z^2$ are based on extrapolations which
still need further confirmations. However in the region of low temperature $T < I / k_B$ and
non-degenerate plasmas our rather simple formulae give a rather good behavior
and describe well the transition from low density to the valley of bound states.
We note however that several physical effects as e.g. plasma phase transitions \cite{RedBook,EbRi85}
are not yet described by the present approach. Evidently this effect appears only in higher order terms or
after transition to some chemical picture \cite{RedBook,EbRi85}. However this is not our
aim here.

In order to describe the transition to full ionization in the degenerate region
as well,
additional effects have to be taken into account, in particular the symmetry
between
electrons and protons is lost. The disappearance of the bound states with
increasing density was discussed in \cite{EbBlRRR09,EbKrRo11}; for more
information see \cite{GreenBook, KSKBook}.

\section{Discussion and Conclusion}

It is shown in this work that transitions from full to partial ionization in the temperature range $10^4 - 10^5$ K
which are due to the formation of H-atoms, are well described in the physical picture. Ring and ladder diagrams
in higher order approximations as well as certain combinations of these basic diagrams are taken into account.
In particular the divergence of the partition function of the Bohr atom
and the state of art concerning this problem is discussed.

Formally the divergence of all partition functions within the frame of the Rutherford model is due
to the infinite range of the Coulomb potential. Practically, a divergent partition function means
that ionization is impossible, atoms are absolutely stable. As stated already by pioneers like
Herzfeld, Fermi, Planck and others, this makes no sense and there should be physical reasons which
make the atomic partition function finite. Among the physical effects leading to a cut of the partition function at some level,
which were mentioned already by the pioneers, are:\\
(i) The high levels $|E_s| < k_BT$ are not stable against collisions in a thermal system and are strongly perturbed by
screening effects, this is represented by the BPL reduction. \\
(ii) At higher densities and moderate temperatures, the contribution of the
partition function provided by the ladder contributions strongly increases, also the nonideality strongly increases. The total contribution
of screening and bound states remains however finite since the
fugacities of the charges become small. The nonideality
contributions saturate and lead to a reduction of the
pressure to about one half of the ideal pressure (see Fig. 4).\\
(iii) At very high densities beyond $10^{23} $ cm$^{-3}$ the electron
degeneracy delocalizes the electrons and makes atoms obsolete.
The nonideality decreases again, the plasma leaves the region of nonideality (screening and
bound states).

\begin{figure}
\centering{\epsfig{figure=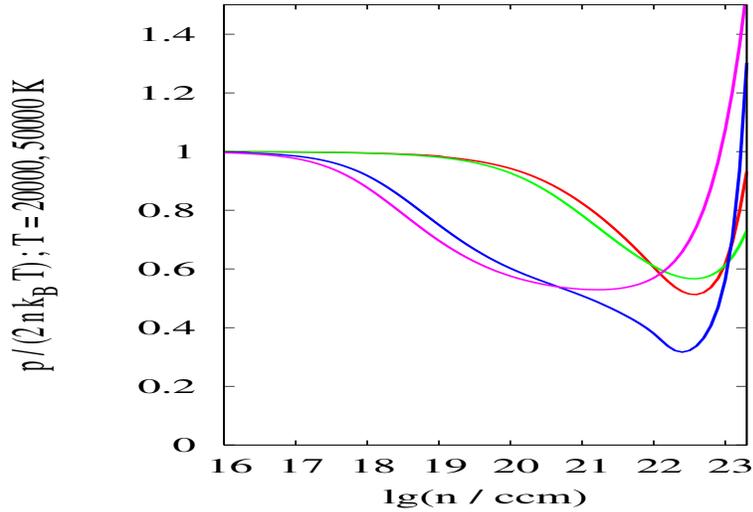,height=7cm,width=0.75 \textwidth,angle=0}}
\caption{ \label{VglNewOld}
Comparison of the analytical theory  for $T = 20 000$ K (magenta) and $T = 50
000$ K (green) to the results of a  numerical code developed earlier within the
chemical picture \cite{befjnrr,befjrr99,EbHi02,befjrr01,
Gericke11} (blue:
$T = 20 000$ K, red: $T=50000$ K). We show the relation
of the pressure to the classical ideal pressure of the ions $n k_B T$.
The largest deviations between the curves for $20000$ K are due to the formation of H$_2$ molecules (deep minimum
in the blue curve) which are not taken into account in the present analytical theory.
}
\end{figure}

\begin{figure}
\centering{\epsfig{figure=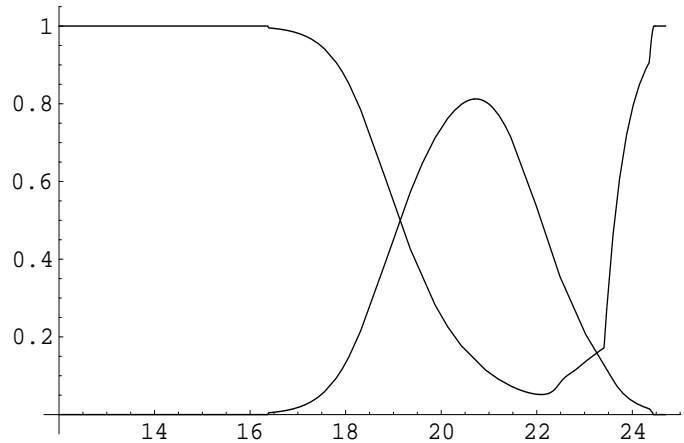,height=6cm,width=0.6 \textwidth,angle=0}}\\
\caption{ \label{IonDiss}
Typical isotherms of the degree of ionization at dissociation of hydrogen for $T
= 20 000 K$ in a wide density range based on the code developed earlier within
the chemical picture \cite{befjnrr,befjrr99,EbHi02,befjrr01}. We show  the
degree of ionization (upper curve) and the degree of protons bound in H$_2$
molecules (transient maximum) as a function of density in log-scale.
}
\end{figure}

The principal scheme how the formation of bound states depends on the density
at a fixed temperature
is shown in Figs. \ref{VglNewOld} and \ref{IonDiss}.
In Fig. \ref{VglNewOld} a comparison of the theory given here for $T = 20000$ K, $T=50000$ K with the
numerical results obtained within an advanced chemical picture by minimization of the free
energy is presented.
With the chemical picture we mean here a description which starts
from an expression for the free energy depending on the density of free charges, atoms and molecules.
One can see that the overall agreement is quite reasonable, the deviations increase only at large densities beyond
$n \simeq 10^{21}$ cm$^{-3}$. The largest deviations appear near the minimum
of the relative pressure for Hydrogen at $20000$ K and
may be interpreted by the formation of molecules which were not taken into account in the present work.
These interpretations are supported by the the last Fig. \ref{IonDiss}. Here
we show the degrees of ionization and the degree of molecular bound states
obtained within the chemical picture for $T = 20000$ K.
The composition and state of the system was obtained by minimization of the free energy.
The numerical results used in Figs. \ref{VglNewOld} and \ref{IonDiss} for comparison were obtained 
with a code developed in some earlier work \cite{EbHi02,MottBook,EbBlRRR09}.


Looking at the pressure curve in Fig. \ref{VglNewOld} one observes as in earlier sections first a decrease corresponding to the formation
of bound states (atoms and molecules). This decrease comes to saturation and we observe a valley of developed bound states.
With increasing density the spatial occupation of the atoms and molecules as well as effects of electronic degeneracy
come into play and the bound states are destroyed. The largest deviations
between the curves for 20000 K are due to the formation of H$_2$ molecules
(deep minimum in the blue curve) which are not taken into account in the present
analytical theory and also effects of strong degeneracy. These rather
complicated effects are not discussed here.
Parallel to the destruction of the bound states we observe an increase of the relative pressure
in the region of rather high densities $n > 10^{22}$ cm$^{-3}$. The monotonic increase of the relative pressure
may be interrupted at temperatures below $10000$ K by a small wiggle. This is not an artifact. A rather flat region of the pressure
in dependence on the densities is contained. At lower temperatures the wiggle may be even stronger
and should be discussed in connection with the possible
existence of plasma phase transitions \cite{RedBook,EbRi85,Chabrier,befjnrr,befjrr99,
RedmerReport,befjrr01}. Recent work in this area was reported in
\cite{Lorenzen,Morales} and found a phase transition.
This is not our topic here, the purpose of the present work is to consider the region of lower densities and higher temperatures.
Our aim is, to obtain similar results as shown in Fig. \ref{VglNewOld} without referring to the rather complex
chemical description based on atoms and molecules and their interactions. Instead of the many assumptions contained
in any chemical description, the calculations given here are based only on Rutherfords model of point charges and
Coulomb interactions as well as on the exact principles of quantum statistics. No a priori knowledge about the properties of atoms
and molecules is required. The price to pay is that we have to avoid the regions where these species are dominant.
With respect to accuracy the rather simple formulae given here may not compete with the recent successful
approaches to dense plasmas based on numerical methods \cite{MilCep01,Gericke11,Lorenzen}. We are convinced however, that
in particular to experimentalists, analytical approaches may always be quite useful in complementing numerical results.
In conclusion, we
first analyzed the basic results for the quantum statistics of the Rutherford model of matter, partially in the historical context starting with the work of Bohr, Herzfeld, Saha, Planck, and Brillouin.
Then we discussed the essential results of quantum statistics since the 60th of the 20th century.
The main new results obtained in this work are:
For Hydrogen plasmas at lower density the EOS within the physical picture was obtained using a combined density fugacity representation which contains the essential contributions describing Saha effects and nonideality. The good convergence of the fugacity series is explained by the fact that the fugacities of the electrons and the protons are rather small in the bound state region. The obtained new expression for the EOS is valid for non-degenerate plasmas and temperatures between $20\,000 $ {\rm K}  and $100\,000 $ {\rm K}.
Partial ionization, i.e., the appearance of atoms, is restricted to a valley in
the density-temperature plane.
There is a transition from below starting in the region of nondegenerate
electrons and a transition from above beginning in the region of degenerate
electrons. Both transitions are connected to different physical
effects.

I has been shown that in the non-degenerate region, the most essential effects is the
BPL-reduction, resulting in the combined action of collisions and screening by absorbing the orders in $e^2,e^4$ and the compensation effects of the states below and above the series limit.
The most essential new contributions given here are the nonlinear ring and ladder functions introduced into the EOS.
These terms stemming essentially from the grand canonical representations are responsible for a good representation of the EOS
in a wide range of densities and temperatures which essentially covers the range
$20\,000\,\, {\rm K}  < T < 100\, 000$ K  and $ n < 10^{22}$ cm$^{-3}$.
As shown in \cite{EbKrRo11} at high density the Pauli-Fock
contributions are essential for the limitation of the number of energy
levels and determine the transition from full ionization at very high densities
to partial ionization. For a detailed discussion of this problem see,
e.g., \cite{GreenBook,KSKBook}.

\end{document}